\documentclass[reprint,amsmath,amssymb,aps,prl]{revtex4-2}
\usepackage{xcolor}
\usepackage{graphicx}% Include figure files
\usepackage{dcolumn}% Align table columns on decimal point
\usepackage{bm}% bold math

\newcommand\Alfven{Alfv\'en } %Proper Names
\begin{document}

%\preprint{APS/123-QED}

\title{Impact of Two-Population $\alpha$-particle Distributions on Plasma Stability}

\author{Mihailo M. Martinovi\'c}
\email{mmartinovic@arizona.edu}
\affiliation{Lunar and Planetary Laboratory, University of Arizona, Tucson, AZ 85721, USA.}
  
\author{Kristopher G. Klein}
\affiliation{Lunar and Planetary Laboratory, University of Arizona, Tucson, AZ 85721, USA.}

\author{Rossana De Marco}
\affiliation{INAF – Istituto di Astrofisica e Planetologia Spaziali, Via Fosso del Cavaliere 100, 00133 Rome, Italy} 

\author{Daniel Verscharen}
\affiliation{Mullard Space Science Laboratory, University College London, Dorking, RH5 6NT, UK} 

\author{Raffaella D'Amicis}
\affiliation{National Institute for Astrophysics, Institute for Space Astrophysics and Planetology, Rome, Italy} 

\author{Roberto Bruno}
\affiliation{INAF – Istituto di Astrofisica e Planetologia Spaziali, Via Fosso del Cavaliere 100, 00133 Rome, Italy} 

\date{\today}

\begin{abstract}
The stability of weakly collisional plasmas is well represented by linear theory, and the generated waves play an essential role in the thermodynamics of these systems. 
The velocity distribution functions (VDF) characterizing kinetic particle behavior are commonly represented as a sum of anisotropic bi-Maxwellians. 
For the majority of in situ observations of solar wind plasmas enabled by heliospheric missions, a three bi-Maxwellian model is commonly applied for the ions, assuming that the VDF consists of a proton core, proton beam, and a single He ($\alpha$) particle population, each with their own density, bulk velocity, and anisotropic temperature. 
Resolving an $\alpha$-beam component was generally not possible due to instrumental limitations. 
The Solar Orbiter Solar Wind Analyser Proton and Alpha Sensor (SWA PAS) resolves velocity space with sufficient coverage and accuracy to routinely characterize secondary $\alpha$ populations consistently. 
%The instrument also provides VDF measurements at up to every 1 second - comparable to the time over which the proton and alpha gyro-radii are convected by the spacecraft location. 
This design makes the SWA PAS dataset ideal for examining effects of the $\alpha$-particle beam on the plasma's kinetic stability. 
We test the wave signatures observed in the magnetic field power spectrum at ion scales and compare them to the predictions from linear plasma theory, Doppler-shifted into the spacecraft reference frame. 
We find that taking into account the $\alpha$-particle beam component is necessary to predict the coherent wave signatures in the observed power spectra, emphasizing the importance of separating the $\alpha$-particle populations as is traditionally done for protons. 
Moreover, we demonstrate that the drifts of beam components are responsible for the majority of the modes that propagate in oblique direction to the magnetic field, while their temperature anisotropies are the primary source of parallel Fast Magnetosonic Modes in the solar wind. 

\end{abstract}

\maketitle

%\tableofcontents

\section{Introduction}
\label{sec:intro}

Solar wind plasma is weakly collisional, allowing the medium to depart from Local Thermodynamic Equilibrium \cite{Marsch_2012_SSRv}. 
The particle dynamics are described by a Velocity Distribution Function (VDF), which is commonly approximated as a sum of ``component'' VDFs, where each component is associated with a particular particle species (e.g., electrons, protons or helium ($\alpha$) particles) \cite{Marsch_1982}.
In the solar wind, where we have decades of in situ observations \cite{Verscharen_2019_LRSP}, we also commonly observe secondary populations within the VDF of a single particle species, typically referred to as \emph{beams} for ions \cite{Alterman_2018_ApJ,Bruno_2024_ApJ} or \emph{strahl} for electrons \cite{Maksimovic_2005}. 
Each component, when separated from the core component, has distinct densities $n_j$ and can be anisotropic, with distinct temperatures perpendicular and parallel to the direction of the magnetic field $T_{\perp,j},T_{\parallel,j}$ \cite{Matteini_2007_GRL}, or streaming with different bulk velocities, producing a commonly observed inter-component drift \cite{Durovcova_2019_SoPh}.
These anisotropies and relative drifts often store a significant percentage of the system's ``free'' energy \cite{Fowler_1968_AdPlP,Chen_2016_ApJ,Klein_2018,Martinovic_2023_ApJ_Ins_2}.

When the amount of free energy exceeds a threshold value, the plasma emits fluctuations in one or more of the system's normal modes, where each emitted mode acts to reduce the free energy source that drives the emission, bringing the system back towards a state of marginal stability. 
In the case of resonant instabilities, the fluctuations of the electromagnetic field can resonate with a subset of particles, scattering them in velocity space and injecting or removing particle kinetic energy \cite{Gary_1993,Verscharen_2019_LRSP}. 
The resulting fluctuations are ubiquitous in the solar wind \cite{Gary_2016_JGRA,Liu_2023_ApJ} and are commonly observed alongside the out-of-equilibrium VDFs \cite{Martinovic_2021_ApJ_Ins_1}. 
Signatures in time series of electric \cite{Mozer_2021_ApJ} and magnetic \cite{Vech_2021_AA} field, as well as in the Power Spectra (PS) of both quantities \cite{Malaspina_2020_ApJS..246...21M,Bowen_2020_ApJS},  provide estimates of wave parameters that can be directly compared to theoretical models. 
The waves directly resonate with the particles at given resonant velocities \cite{Vech_2021_AA,Shankarappa_2024_ApJ}, emphasizing their crucial yet not sufficiently explored influence on solar wind evolution. 

The interplay of unstable modes and solar wind heating is a central question that needs to be answered for the community to understand the quantitative effects of various solar wind heating mechanisms \cite{Matteini_2012_SSRv}. 
The linear plasma response associated with a VDF is ideally determined by directly integrating functions of the velocity gradients of the phase space density \cite{Verscharen_2018_JPlPh}; this formalism is well known and given in plasma textbooks \cite{Stix_1992,Gary_1993}. 
This calculation is computationally expensive and not suitable for processing mission surveys with millions of observations. 
Instead, a common approach involves the modeling of the VDF as a known function, e.g. a bi-Maxwellian or $\kappa$ distribution, significantly simplifying the integration. 
For any of these VDF models, the resulting plasma response is function of a finite number of VDF parameters, e.g. the density, bulk velocity and temperature anisotropy for each component, further on referred to as elements of the set $\mathcal{P}$. 
For a set of VDF parameters, a plethora of available dispersion solvers \cite{Roennmark_1982,Astfalk_2015_JGRA,Klein_2015_PhPl,Klein_2017_JGRA,Verscharen_2018_RNAAS} and Machine Learning (ML) algorithms \cite{Martinovic_2023_ApJ_Ins_2}, are available to determine the supported normal modes. 

The major stumbling block for an accurate description of the waves supported by the plasma is accurate representation of the VDF. 
In situ particle measurements suffer from inherent limitations in cadence, energy and angular coverage, resolution, and statistical uncertainties \cite{Wilson_2022_FrASS}. 
Even subtle inaccuracies in the VDF representation lead to notable difference in the results of stability analysis \cite{Walters_2023_ApJ}. 
Another fundamental problem in the treatment of measured VDFs appears in the identification and treatment of secondary components. 
In this paper, we focus on ion distributions that are most often described by a single $\alpha$-particle and a single proton Maxwellian components \cite{Kasper_2006_JGRA}. 
More sophisticated treatments of the VDF presume the capability of identifying proton beams, where traditionally the largest difficulty is that, in the raw instrument data, the beam drifts and $\alpha$-particle different mass-to-charge ratio often cause the two populations to overlap \cite{Marsch_1982,Scudder_2015_ApJ,Alterman_2018_ApJ,Durovcova_2019_SoPh}. 
Recent in situ instruments, such as the Solar Probe ANalyzer—Ions (SPAN-I) on \emph{Parker Solar Probe (PSP)}  \cite{Livi_2022_ApJ_SPAN-I} and the Solar Wind Analyser Proton and Alpha Sensor (SWA-PAS) \cite{Owen_2020_A&A_SWA} on \emph{Solar Orbiter (SolO)}  provide significantly increased instrument capabilities in accuracy and sampled particle energy-per-charge range, providing the possibility of fitting the secondary $\alpha$ component. 

In this paper, we explore the effects of $\alpha$-particle beams on plasma stability, comparing the inferred unstable modes from VDF parameters obtained by two models: 
i) 5-component---2-proton, 2-$\alpha$, and 1-electron, and 
ii) 4-component---2-proton, 1-$\alpha$, and 1-electron plasma. 
The $\mathcal{P}$ sets obtained from the two approaches can fundamentally alter the stability analysis. 
For example, a moderate, stable drift with respect to the proton core for a one-component $\alpha$-particle distribution is alternatively represented as small drift of the $\alpha$ core and potentially unstable drift of the $\alpha$ beam. 
Additionally, the single $\alpha$ VDF elongated in the parallel direction is separated into two anisotropic VDFs with $T_{\perp,j}>T_{\parallel,j}$---resulting in distributions with notably different stability properties. 

Observational evidence for the occurrence of the relevant waves arises from the comparison of the inferred modes with the trace PS of the magnetic field, where waves are recognized by a narrow-band coherent signal distinct from broadband background turbulence fluctuations. 
The signature of polarized fluctuations provides confident information about two major parameters: the real part of the wave frequency and wave polarization as measured in the spacecraft frame of reference. 
Using our stability analysis discussed in the following section and accounting for the Doppler shift from the plasma to the spacecraft frame, we directly compare our dispersion-relation solver results with PS observations. 
Previous analyses use similar methods to characterize Ion Cyclotron (IC) heating \cite{Bowen_2020_ApJS,Shankarappa_2024_ApJ} and to describe the impacts of $\alpha$-beams in \emph{PSP} observations \cite{McManus_2024_ApJ}. 
Here, we build on these previous efforts by separating the effects of the two $\alpha$-particle components, quantifying the differences between the two VDF models and illustrating the importance of careful treatment of secondary populations in solar wind kinetic physics. 

\section{Dataset and Methodology}
\label{sec:methods} 

\subsection{Fits of the Velocity Distribution Parameters}
\label{ssec:VDF_fit}

SWA PAS \cite{Owen_2020_A&A_SWA} is an electrostatic analyzer with an energy range sufficient to measure protons with velocities up to $\sim$1500 km/s.
This range is decreased by a factor of $\sqrt{2}$ for $\alpha$-particles due to the difference in mass-to-charge ratio.
SWA PAS has a standard energy resolution of $\sim 9.5\%$, with a 11x9 grid in azimuth and elevation look directions covering a 60x40$^\circ$ range, ensuring the necessary velocity space coverage to resolve both proton and $\alpha$ secondary populations. 

\begin{figure}
\includegraphics[width=0.48\textwidth]{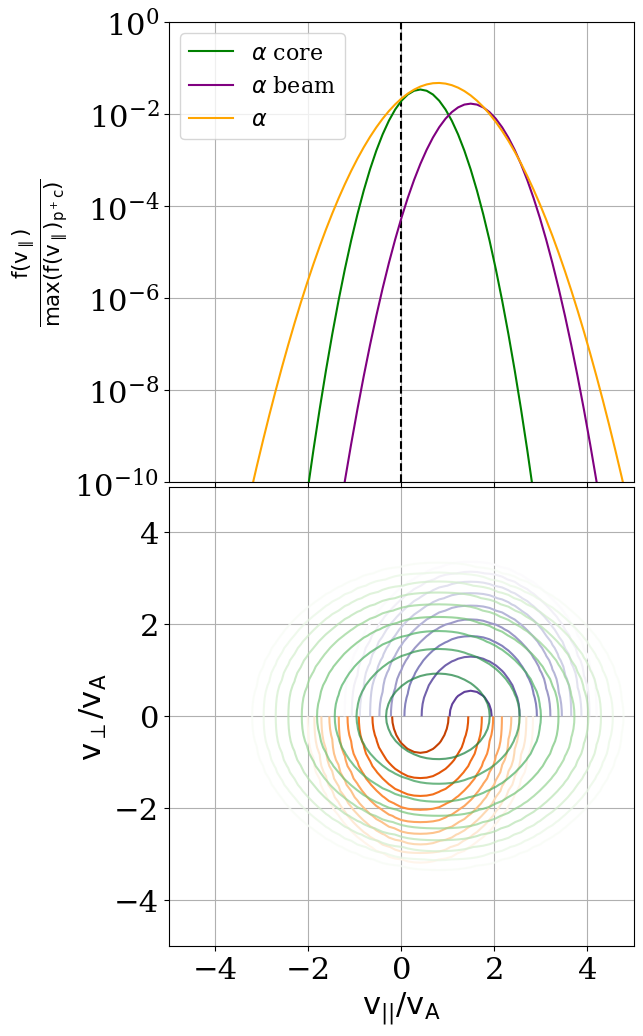}% Here is how to import EPS art
\caption{Comparison of the fits of $\alpha$ components for the VDF in parallel sampled by \emph{SolO} PAS on March 2, 2022 16:18:13 using 4-component (orange) and 5-component (green, purple) model. 
%Parallel cuts at $v_\perp=0$ and 2D phase space contour maps are given in the \emph{top} and \emph{bottom} panels. 
The VDFs are centered with respect to proton core population (not shown), with parameters for both cases listed in Table \ref{tab:VDF}.
}
\label{fig:VDF}
\end{figure}

We obtain the fits for the 4- and 5-component models using the novel technique detailed in \cite{DeMarco_2023_A&A,Bruno_2024_ApJ} applied directly to the full three-dimensional VDFs collected by PAS. 
This approach  leverages the cluster analysis technique commonly used in machine learning (ML) to separate the proton core, proton beam, $\alpha$ core and $\alpha$ beam by identifying sub-components within the VDF, rather than relying on the conventional bi-Maxwellian fit.
Every component of the VDF is given in the form that introduces temperature anisotropies and drifts as
\begin{equation}
    f_j(\mathbf{v}) = \frac{n_j}{\pi^{3/2}w_{\perp j}^2 w_{\parallel j}^2} e^{\frac{-v_\perp^2}{w_{\perp j}^2} - {\frac{(-v_\parallel - \Delta v_\parallel j)^2 }{w_{\parallel j}^2}}}
\end{equation}
where $w_j = \sqrt{2 k_b T_j / m_j}$ and $k_b$ is Boltzmann constant. 
The differences between the two models are illustrated for a representative interval in Figure \ref{fig:VDF} and Table \ref{tab:VDF}. 
Fitting $\alpha$ particles as one or two components renders significant differences in the obtained $\mathcal{P}$ sets and potential sources of free energy. 

\begin{table}
    \centering
    \begin{tabular}{|l|l|l|}
         \hline
        $n_{\alpha}/n_{pc}$ = 4.8\% & $n_{\alpha c}/n_{pc}$ = 3.4\% & $n_{\alpha b}/n_{pc}$ = 1.7\% \\
         \hline
        $T_{||,pc} / T_{||,\alpha}$ = 0.11 & $T_{||,pc} / T_{||,\alpha c}$ = 0.27 & $T_{||,pc} / T_{||,\alpha c}$ = 0.20 \\
         \hline
        $T_{\perp,\alpha c} / T_{||,\alpha}$ = 0.70 & $T_{\perp,\alpha c} / T_{||,\alpha c}$ = 1.73 & $T_{\perp,\alpha b} / T_{||,\alpha b}$ = 1.52 \\
         \hline
        $\Delta v_{\alpha c}/v_A$ = 0.78 & $\Delta v_{\alpha c}/v_A$ = 0.40 & $\Delta v_{\alpha b}/v_A$ = 1.48 \\
         \hline
    \end{tabular}
    \caption{Parameters of the VDFs shown on Figure \ref{fig:VDF}.
    }
    \label{tab:VDF}
\end{table}

\begin{figure*}
\includegraphics[width=0.8\textwidth]{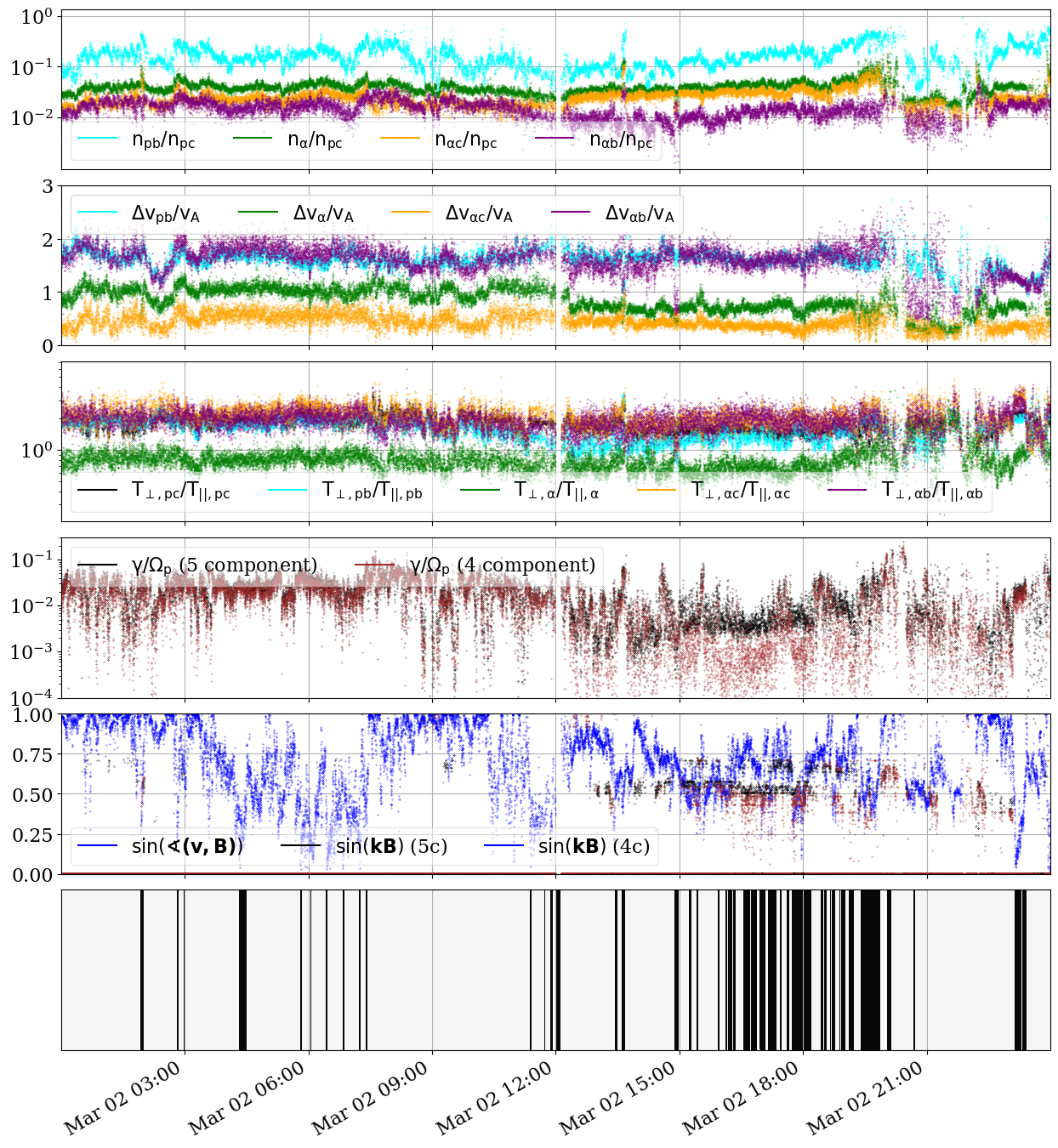}% Here is how to import EPS art
\caption{ Overview of the VDF parameters used for stability analysis on March 02, 2022. 
The \emph{top three} panels show the density, drift, and temperature anisotropy for proton core (black), proton beam (cyan), $\alpha$ particles as a single population (green), $\alpha$ core (brown), and $\alpha$ beam (purple). 
The normalized growth rates and directions of Most Unstable Modes with respect to the magnetic field and solar wind speed in the spacecraft frame for both models is given in the \emph{fourth} and \emph{fifth} panels. 
The intervals where the measured magnetic field polarization in the spacecraft frame marks a clear presence of coherent waves---$\sigma > 0.6$ or $\sigma < -0.6$ (Equation \ref{eq:MVA})---are marked in black in the \emph{bottom} panel. 
}
\label{fig:overview}
\end{figure*}

We analyze observations from one representative day, March 02, 2022. 
The observations are available at $\sim$4s cadence, leading to 16,655 successfully processed intervals for both models, which are used for stability analysis. 
The quantities of major free energy sources are provided in the \emph{top three} panels of Figure \ref{fig:overview}. 
These panels are clear examples of the common scenario shown in Figure \ref{fig:VDF}---when treated as a single Maxwellian, the drifts of the $\alpha$ particles take values in between the core and beam components from the 5-component fits (\emph{second} panel), while at the same time having artificially increased parallel temperature (\emph{third} panel). 
Moreover, the drifts of proton and $\alpha$ beams seem to maintain very similar values.
Here, the drifts are normalized to the local \Alfven speed $v_A = B/\sqrt{\mu_0 n_{pc} m_p}$, where $\mu_0$ is the magnetic permeability of vacuum and $m_p$ is the proton mass. 

\subsection{Linear Stability Analysis}
\label{ssec:PLUMAGE} 

Traditional solvers provide solutions to the wave equation for every mode supported by the system, namely those that satisfy 
$\mathcal{D} (\omega, \mathbf{k}, \mathcal{P})=0$, where $\mathcal{D}$ is the vector wave equation for a given complex frequency $\omega=\omega_{\textrm{r}} + i \gamma$, wavevector $\mathbf{k}$, and a set of VDF parameters $\mathcal{P}$. 
In this work, we determine the stability of the VDF with respect to all the ion scale modes by employing the \texttt{PLUMAGE} solver \cite{Klein_2017_JGRA}, which calculates the Nyquist stability criterion for a hot magnetized collisionless plasma.
Over the upper half of the complex frequency plane, the space that contains unstable modes with $\gamma > 0$, the solver performs the contour integral
\begin{equation}
    W_n(\mathbf{k}, \mathcal{P}) = \frac{1}{2 \pi i} \oint \frac{\mathrm{d} \omega}{det [\mathcal{D} (\omega, \mathbf{k}, \mathcal{P})]}
\end{equation}
to obtain the count of the unstable modes. 
This integration is repeated after shifting the contour from $\gamma=0$ to $\gamma_\epsilon>0$ until $W_n$ expresses only a single Most Unstable Mode (MUM).

The values of $\omega$ and $\mathbf{k}$ for a MUM provided by \texttt{PLUMAGE} are then inserted into the a \texttt{PLUME} dispersion solver \cite{Klein_2015_PhPl}, finding a detailed set of MUM parameters---further on referred to as $\mathcal{W}$---that contains electromagnetic eigenfunctions $\delta \mathbf{E}, \delta \mathbf{B}$, density and velocity fluctuations for each component $\delta n_j$ and $\delta \mathbf{v}_j$, mode polarization $\sigma$, and estimated emitted or absorbed power by each component over a wave period $P_j$. 
The description of the formalism that calculates the parameter set $\mathcal{W}$ is given elsewhere \cite{Klein_2013_PhD,Huang_2024_JPlPh}. 
Stability analysis that simultaneously uses \texttt{PLUMAGE} and \texttt{PLUME} was previously performed for \emph{Wind} \cite{Klein_2018}, \emph{Helios} \cite{Klein_2019_ApJ,Martinovic_2021_ApJ}, and \emph{PSP} \cite{Klein_2021_ApJ,McManus_2024_ApJ} observations, with additional technical details described in those references. 

The \emph{fourth} and \emph{fifth} panels in Figure \ref{fig:overview} show the normalized growth rates 
$\gamma/\Omega_p$, where $\Omega_p = e_c B / m_p$ is proton cyclotron frequency, and $e_c$ is the elementary charge. 
Linear Vlasov--Maxwell theory is valid for $\gamma/\omega_r \ll 1$. 
Since we observe waves with $\omega_r \sim \Omega_p$, the results are comfortably in the domain well described by linear theory. 
Aside from the time interval 15:00 - 19:30, most of the unstable modes predicted propagate in the directions parallel or antiparallel  to the magnetic field. 

\subsection{Comparison of Inferred Waves with Observed Power Spectra}
\label{ssec:MWT_Doppler}

The magnetic field vector measurements are provided by the \emph{SolO} magnetometer \cite{Horbury_2020_A&A_MAG} operating in its normal mode producing 8 samples per second. 
A Morlet Wavelet Transform (MWT) \cite{Torrence_1998_BAMS} $\mathbf{\widetilde{B}}$ combined with Minimum Variance Analysis (MVA) \cite{Sonnerup_1998_ISSIR} provides both PS amplitudes and spacecraft frame polarization as functions of time and frequency throughout the day. 
In contrast to magnetic helicity, polarization $\sigma$ is defined as the normalized reduced helicity along a single direction as observed by the spacecraft \cite{Matthaeus_1982_JGR}. 
The MWT for each timescale $s\sim f^{-1}$ and time $t$ is transferred to a coordinate system where the average magnetic field is along z direction, and the minimum variance eigenvector component is along x axis. 
Then, we define $\sigma$ as
\begin{equation}
 \sigma = \dfrac{2  \text{Im}\big[\widetilde{B_x}(s,t) \widetilde{B_y^*}(s,t)\big]}{|\widetilde{B_x}(s,t)|^2 + |\widetilde{B_y}(s,t)|^2}
 \label{eq:MVA}
\end{equation}
where star stands for complex conjugate. 
The polarization values provided by MVA are highly variable, and smoothing over time and frequency is required in order to reveal a clear signature of coherent waves on top of the non-coherent background plasma turbulence. 
The polarization values vary from $-1$ for Right-Handed (RH) to $+1$ for Left-Handed (LH) waves, commonly referred to as electron-resonant and ion-resonant fluctuations \cite{Bowen_2020_ApJS,Shankarappa_2024_ApJ}. 
The details of our procedure are given in Appendix B of \cite{Shankarappa_2024_ApJ}. 
For direct comparison between the stability analysis and the MVA results, the real frequency values provided by \texttt{PLUMAGE} must be converted into the spacecraft frame of reference. 
The standard Doppler shift relation $\omega_{sc} = \omega_{\text{plasma}} + \mathbf{k} \cdot \mathbf{v_{\text{sw(sc)}}}$ applies, where $\mathbf{v_{\text{sw(sc)}}}$ is the solar wind speed in the spacecraft frame. 
The procedure requires careful treatment of variables with respect to inferred wave direction, spacecraft speed, and wave polarization. 
Details of the calculation for different types of modes are provided by \cite{Shankarappa_2024_ApJ}. An overview of the results is given in Figure \ref{fig:PSD}. 

Direct detection of unstable waves is not always possible due to effects related to the sampling direction. 
The fluctuations with wavevectors parallel to the velocity vector require lower emitted power in order to be observable in the PS \cite{Woodham_2021_ApJ}. 
Also, fluctuations are  only detected if their spacecraft frame frequency is lower than the 4 Hz Nyquist frequency of the MWT and if they are continuously emitted for at least several $2\pi \omega_r^{-1}$ periods so that their signature is not removed by our smoothing method. 
In practice, this set of conditions suggests that mildly unstable modes are difficult to detect due to polarization fluctuations in the background turbulence, while strongly unstable modes---burst-like waves that are emitted over short times---are  not detected in the PS due to being smoothed out. 
Indeed, the \emph{fifth} and \emph{sixth} panels of Figure \ref{fig:overview} show that the clear PS signatures mostly originate from moderately unstable $ \sim 10^{-3} \lesssim \gamma/\Omega_p \sim 10^{-2}$ modes.

\section{Results and Discussion}
\label{sec:results} 

\subsection{Direct Comparison of Inferred and Measured Instabilities} 
\label{ssec:overview} 

\begin{figure*}
\includegraphics[width=0.75\textwidth]{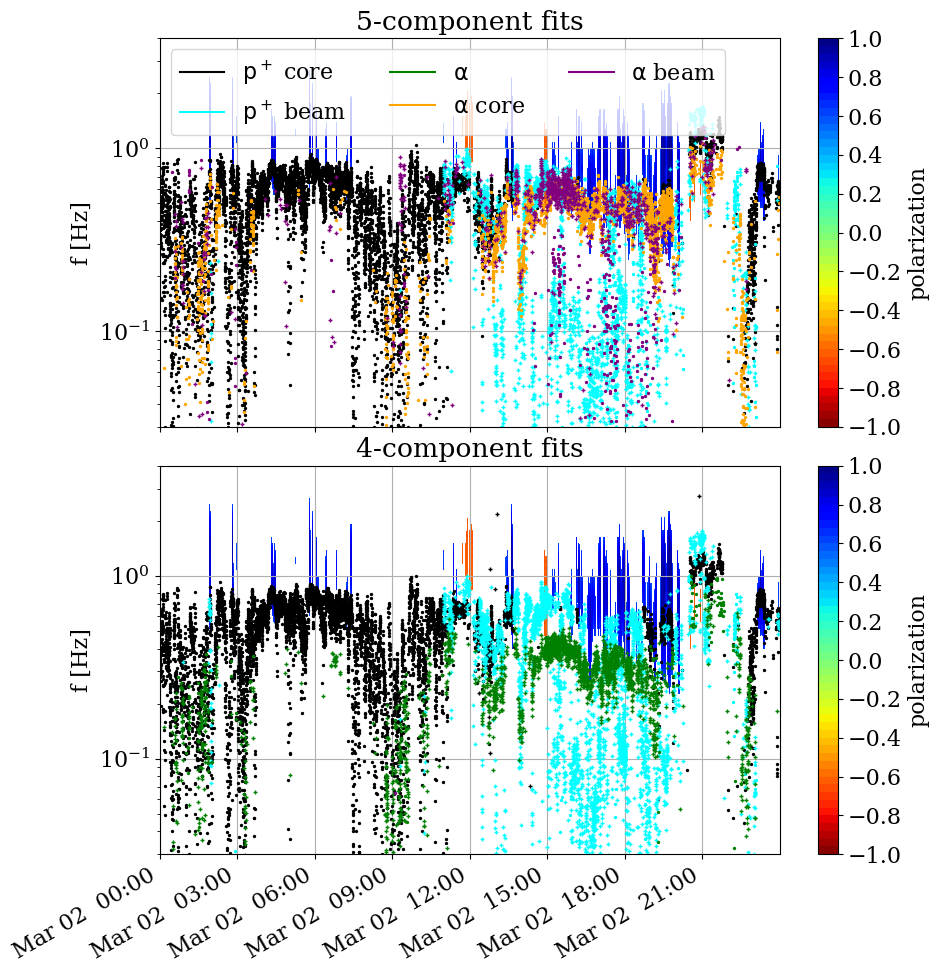}
\caption{Direct comparison of the frequency and $|\sigma| \geq 0.6$ of measured coherent magnetic field fluctuations in the spacecraft frame with the inferred MUM for each interval. 
LH polarized modes are shown with dots and RH with + symbols. 
The color coding of the strongest emitting component for each MUM is the same as in Figure \ref{fig:overview}. 
}
\label{fig:PSD} 
\end{figure*}

\begin{figure}
\includegraphics[width=0.38\textwidth]{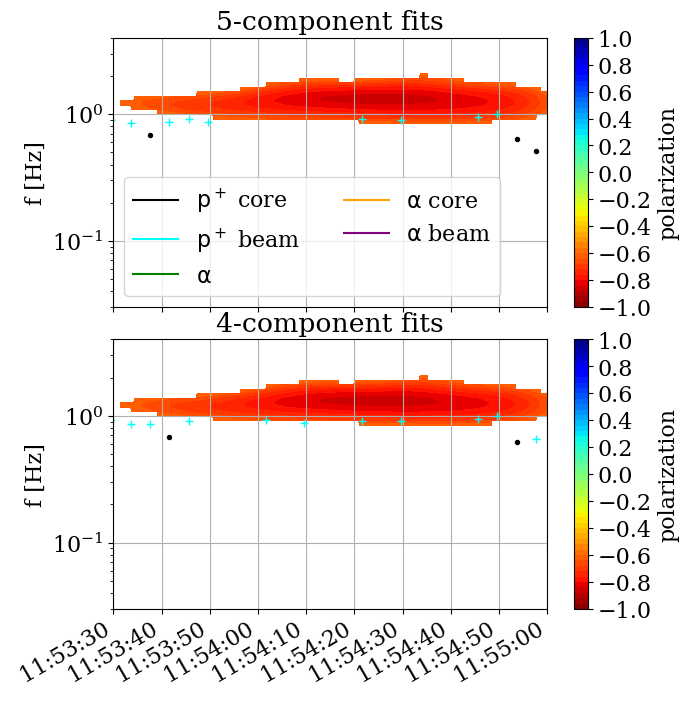}
\includegraphics[width=0.38\textwidth]{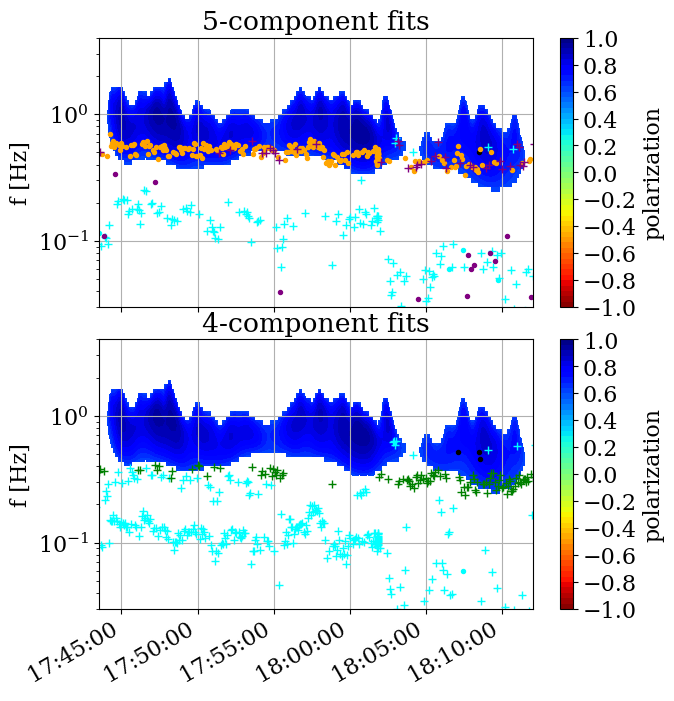}
\includegraphics[width=0.38\textwidth]{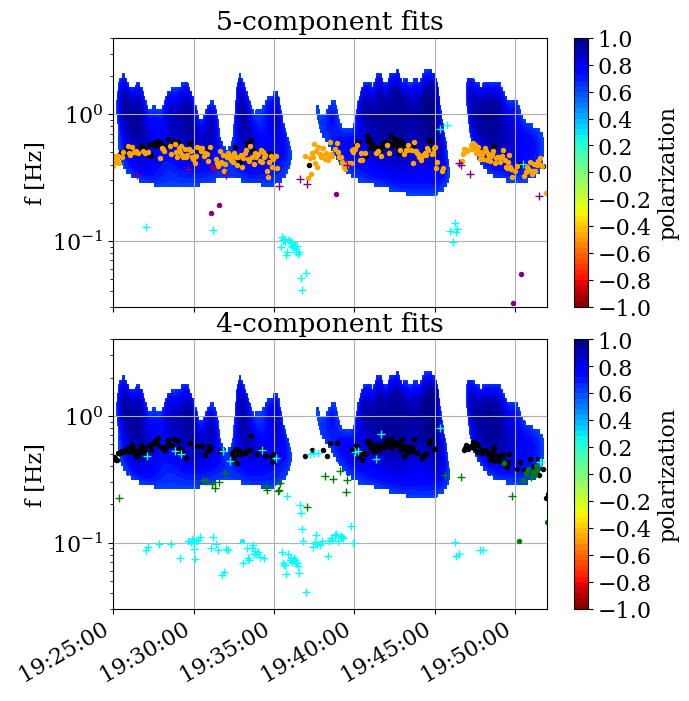}
\caption{Same as Figure \ref{fig:PSD}, with magnification on specific intervals where RH (\emph{top}) and LH (\emph{bottom}) modes are predicted by both models. 
\emph{Middle} panel shows the period where the correctly estimated $\alpha$ core temperature anisotropy is necessary for the correct description of the observed waves. 
}
\label{fig:PSD_cuts} 
\end{figure}

Our results are summarized in Figure \ref{fig:PSD}. 
Coloring represents the strongest emitting component---the VDF component with maximum $P_j$---that drives the MUM for a given interval. 
In the period prior to noon on March 2, the proton core is the strongest emitting component for the vast majority of MUMs. 
Since the proton core components for both models are largely the same, the inferred modes have very similar properties and the stream is dominated with LH IC modes (black dots) characteristic of fast, collisionally young solar wind \cite{Martinovic_2021_ApJ_Ins_1}. 
Similar behavior is recorded for the period between 11:30 and 13:00, when the plasma is primarily influenced by the modes created by the proton beam (Figure \ref{fig:PSD_cuts}). 
As the observed RH polarization is less abundant in our data set, we show the details of one of these intervals in Figure \ref{fig:PSD_cuts}, \emph{top}. 
The analysis of the dispersion relation of both 4-component and 5-component VDF models provides the same RH solutions with proton beams as the emitting population. 

The importance of the more precise treatment of the $\alpha$ VDF is visible during the second half of the day. 
Here, 4-component analysis predicts constant presence of the modes induced by the proton beam and the $\alpha$-particle components in the time period 12:00-19:30. 
However, the inferred frequencies in the spacecraft frame for the $\alpha$-particle induced modes are significantly lower than the ones observed, while the proton-beam-induced modes match with the observed PS in frequency, but not in polarization. 
The example in Figure \ref{fig:PSD_cuts} (\emph{middle}) shows that these intervals are  appropriately addressed by the 5-component approach. 
The observed LH modes are induced by $\alpha$ core anisotropy, which is increased by approximately a factor of 3 for 5-component cases due to the separation of the $\alpha$ beam. 

A case in which both proton and $\alpha$ core components drive the same mode unstable with the same polarization and frequency is shown on Figure \ref{fig:PSD_cuts} (\emph{bottom}). 
Although we cannot distinguish which component of the VDF emits the observed waves from MVA analysis alone, the 5-component prediction for $\alpha$ core induced IC waves shows more consistent emission to accompany the strongly polarized observed PS over a wide band of frequencies. 

Overall, the stability analysis based on 5-component fits shows consistent agreement with the observed PS both in values of observed real frequency and polarization. 
For the 4-component approach, such agreement is observed only for the MUMs driven by the proton core and beam. 
This is an expected result, as the proton parameters are quantitatively similar for both models. 

%The significance of accounting for $\alpha$ beams is improvements is twofold. 
%First, 
These results have a profound consequence on our understanding of in situ heating of $\alpha$ particles. 
Remote sensing observations of the corona suggest that the plasma is highly anisotropic in the perpendicular direction ($T_\perp > T_\parallel$) for both protons and $\alpha$-particles \cite{Cranmer_2002_SSRv}. 
However, the in situ data sampled by \emph{Helios} at 0.3-0.7 au clearly suggest the opposite ($T_{\perp,\alpha} < T_{\parallel,\alpha}$), with notable, and sometimes extreme, elongations of the $\alpha$ population in the parallel direction \cite{Stansby_2018_SolPh,Durovcova_2019_SoPh}. 
Although several mechanisms, such as Landau Damping \cite{Shankarappa_2023_ApJ}, Transit Time Damping \cite{Huang_2024_JPlPh}, or Magnetic Pumping \cite{Lichko_2017_ApJL,Lichko_2020_NatCo,Montag_2022_PhPl} preferably heat particles in the  direction parallel to the magnetic field, none of them are expected to be highly effective towards $\alpha$ particles and not simultaneously affect protons in the same fashion \cite{Cranmer_2012_ApJ}. 
On the contrary, it is well established that the processes that preferably heat the plasma in the perpendicular direction, such as IC heating \cite{Bowen_2020_ApJS,Bowen_2024_ApJ,Afshari_2024_Nature} and Stochastic Heating \cite{Chandran_2010,Martinovic_2020_ApJS} are not just measured throughout the inner heliosphere, but also more prominent at lower radial distances \cite{Martinovic_2019_ApJ,Liu_2023_ApJ}. 

Observations from both \emph{Helios} and \emph{ACE} reveal that the temperature ratio between $\alpha$ particles and protons typically peaks around 1 in the slow solar wind, consistent with the expectations for an isothermal fluid \citep{Kasper_2008}. 
In contrast, this ratio peaks at 4 in the fast solar wind, signaling equal thermal speeds, where heating processes during wind expansion preferentially heat the minority species \citep{Tu_2001_JGR,Marsch_2001_JGR,Kasper_2008}. 
In some cases, the temperature ratio exceeds 5, indicating the possible involvement of an additional anomalous heating mechanism \citep{Kasper_2008}. 
However, while such mechanisms cannot be entirely ruled out, \cite{Bruno_2024_ApJ} recently showed that separating the $\alpha$ beam from its core population restores the expected value of 4 when focusing solely on the temperature ratio between the $\alpha$ core and the proton core. 
These authors suggest that the apparent anomalous increase in $\alpha$ particle temperatures can instead be attributed to the presence of the relatively massive $\alpha$ beam drifting along the local magnetic field.

\subsection{Analysis of Mode Characteristics}
\label{ssec:stats} 

\begin{figure*}
\includegraphics[width=0.75\textwidth]{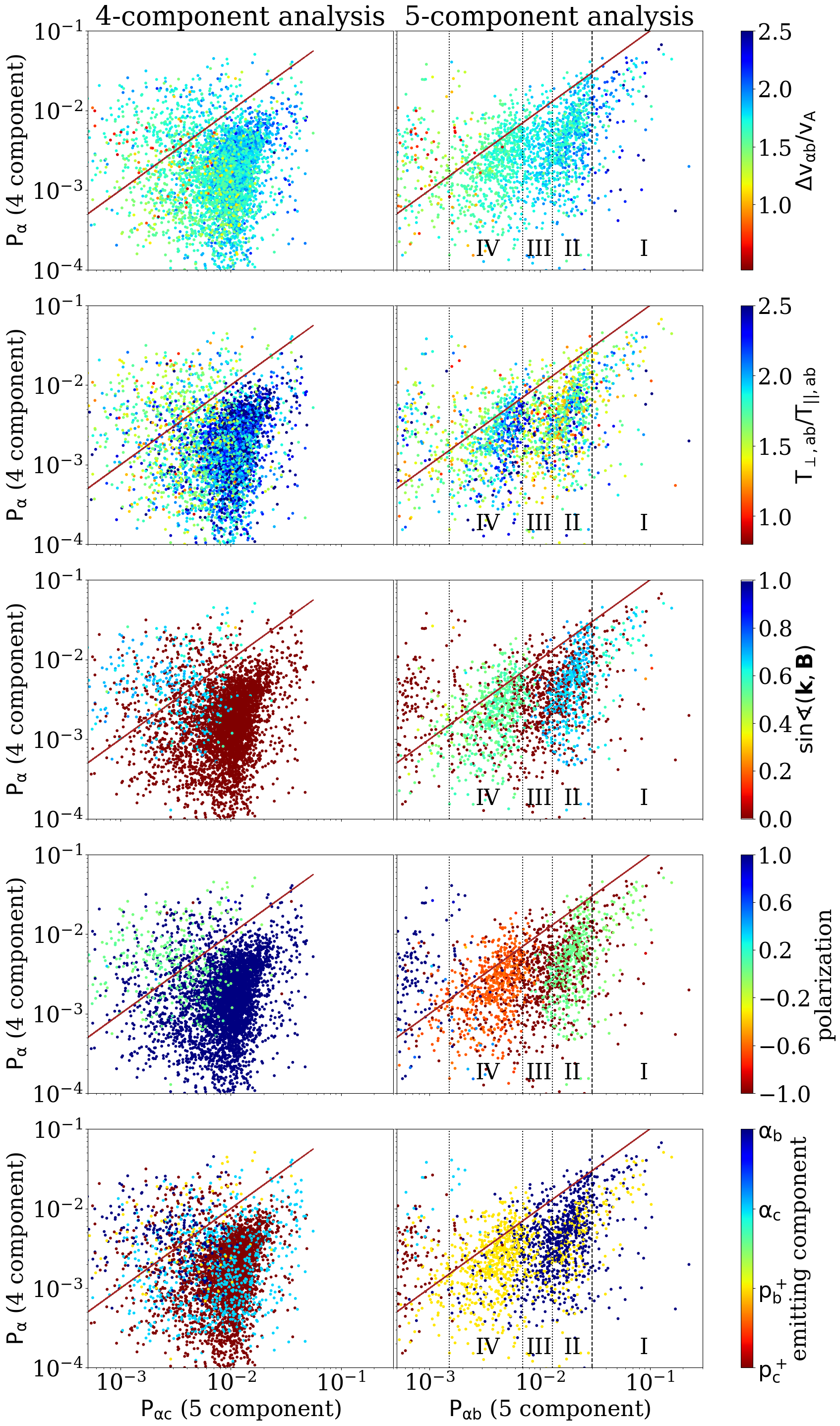}
\caption{Comparison of the power emitted by a single $\alpha$ component (4-component) and $\alpha$ core and beams (5-component analysis), depending on: $\alpha$ beam drift, $\alpha$ beam temperature anisotropy, MUM propagation angle, polarization, and emitting component. 
Brown lines mark equal emitting powers. 
Areas marked with Roman numbers are discussed in the text. 
}
\label{fig:stats} 
\end{figure*}

\begin{figure}
\includegraphics[width=0.42\textwidth]{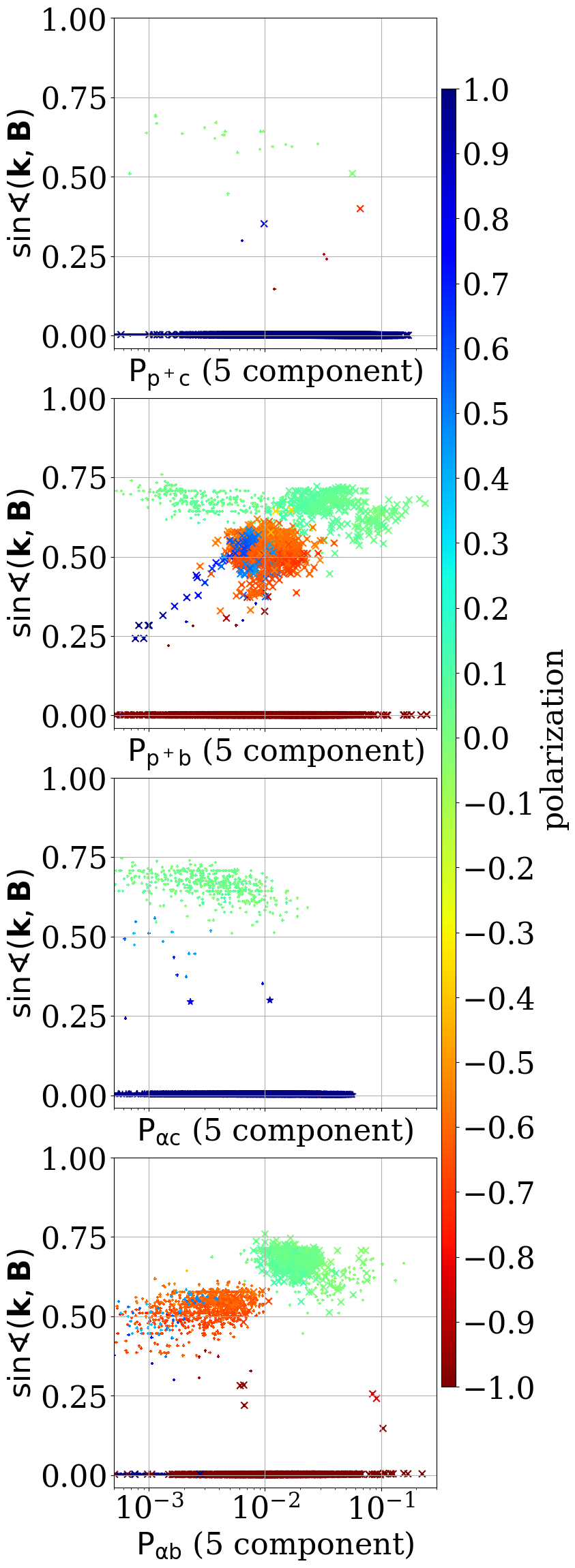}
\caption{Emitted power over one wave period for each of the VDF components. 
Intervals in which the given component drives the MUM are marked by X, while others are shown with dots. }
\label{fig:angles} 
\end{figure}

We use the change in polarization $\sigma$ of MUMs and its propagation direction with respect to the magnetic field to illustrate the behavior of different mode types. 
After testing all the elements of $\mathcal{P}$ related to the $\alpha$ beam component, we find that the drift $\Delta v_{\alpha b}/ v_A$ is the only feature that clearly correlates with $P_{\alpha b}$. 
Figure \ref{fig:stats} (\emph{top right}) illustrates the increase of the emitted power with increasing drift speed---a relation that cannot be inferred from the 4-component treatment. 
Other panels reveal the transition between unstable modes as $\Delta v_{\alpha b}/ v_A$ changes. 
When the $\alpha$ beam drift is strong (above $\sim 2v_A$), it induces the parallel Fast Magnetosonic Mode (FMM) as the MUM of the system (area I on Figure \ref{fig:stats}). 
This is the same mode that becomes unstable for high $\Delta v_{pb}/ v_A$ \cite{Daughton_1998_JGR}. Although the proton beam drift has similar values to the $\alpha$ drift for the majority of our intervals, the MUM obtained by \texttt{PLUMAGE} analysis is primarily powered by the $\alpha$ drift. 
These cases require both $\Delta v_{\alpha b}/ v_A$ and $T_{\perp,\alpha b} > T_{\parallel,\alpha b}$ to be substantially large and are rare for the heliocentric distances covered by \emph{SolO} observations. 
Further exploration via 5-component treatment of \emph{PSP} VDF data is required for further understanding of these modes. 

As $\Delta v_{\alpha b}/ v_A$ decreases, we still find the $\alpha$ beam to be the primary emitting component, but with the drift as the primary source of free energy. 
Characteristic oblique FMMs appear, acting to decrease the drift, while partially dispersing free energy towards increasing the $\alpha$ beam anisotropy (II). 
This part of parameter space maintains analogy between behavior of proton and $\alpha$ beams, as similar FMM-induced perpendicular diffusion of protons is predicted in simulations \cite{Ofman_2022_ApJ,Pezzini_2024_arXiv} and through quasi-linear theory \cite{Shaaban_2024_arXiv}, again leading to conditions for the appearance of parallel FMMs induced by the $\alpha$ beam anisotropy. 
A specific profile of $T_{\perp,\alpha b} > T_{\parallel,\alpha b} (\Delta v_{\alpha b}/ v_A)$ on \emph{second right} panel of Figure \ref{fig:stats} shows the temperature anisotropy decreasing as the drift decreases, just to sharply rise as the drift is further decreased by the oblique FM. 
However, the increased $T_{\perp,\alpha b} > T_{\parallel,\alpha b}$ for lower drift emits significantly less power (\emph{top right}) and is no longer sufficient to be classified as the MUM of the system. 

The oblique fast mode, primarily induced by the proton beam drift, becomes the MUM (IV) once $\Delta v_{\alpha b}/ v_A$ drift speed is low enough. 
This ``background'' mode is often responsible for maintaining the proton beam drift within the confines of marginal linear stability \cite{Martinovic_2023_ApJ_Ins_2}. 
Here, a similar type of oblique FM instability (II) is most likely responsible for limiting the values of $\Delta v_{\alpha b}/ v_A$. 
We emphasize that the two oblique FMMs (II and IV), although induced by similar types of free energy sources, have different effects on the plasma dynamics. 
The panels in \emph{third} row of Figure \ref{fig:stats} and Figure \ref{fig:angles} demonstrate that, due to different propagation angles, the two modes can have fundamentally different polarizations. 
Depending on the combination of $\mathcal{P}$ parameters, most importantly the normalized drift, the unstable modes can be either linearly or circularly polarized. 
Both proton and $\alpha$-particle beams feature both spherical and linear polarizations, although we do not observe $\alpha$ beam RH oblique modes to be MUMs, probably because these modes appear at lower $\Delta v_{\alpha b}/ v_A$ and therefore lower $ P_{\alpha b}$ values. 
%This property alone is an important argument for the separate treatment of the secondary $\alpha$ component, as it leads to different resonances and fundamentally different effects on the VDF. 

These oblique modes are neither induced by proton nor $\alpha$ core populations. 
Moreover, there is a clear distinction between the core populations consistently causing LH, IC mode instability, with some exceptions for the power emitted in oblique directions of propagation  by the $\alpha$ core (\emph{third} panel), which match well with the cases of increased $\Delta v_{\alpha c}/ v_A$ drift (not shown). 
Parallel-propagating modes are also driven by the beams, with the beams constantly emitting FMMs instead of IC modes. 

Our division of power emission into components neglects the effects of the energy exchange between the components at the resonant velocities specific for a given mode's frequency $\omega_r$ and wavevector $\mathrm{k}$. 
These interactions can lead to one or more components partially absorbing the power emitted by another component, leading to subtle changes in the growth rate and eigenfunction values. 
It would be worthwhile to  investigate the intricacies of these particle--mode resonances in future work. 

\section{Conclusions}
\label{sec:conclusions} 

The results presented here can be summarized by two important points. 
In general: 1) while parallel-propagating modes are primarily induced by anisotropies, oblique modes are primarily induced by the drift, and 2) core populations are primarily responsible for parallel LH IC modes, while beams are primarily responsible for parallel RH fast modes. 
Although these rules have exceptions that we will elaborate in detail in future work, this highly simplified description is surprisingly accurate for the description of solar wind stability dynamics. 

The behavior of MUMs induced by proton and $\alpha$-particle beams show striking similarities. 
Neglecting both the proton beams \cite{Klein_2021_ApJ} or $\alpha$ beams a from the linear stability analysis removes the prediction of oblique modes and RH FMMs that are clearly observed in the magnetic field PS. 
Also, both components feature an oblique FM at lower drifts, which is responsible for maintaining similar drift values and trends for the two components. 

The implications of the difference in treatment of the VDF have wider implications than just changes of the inferred instabilities. 
Depending on the 4-component or the 5-component model, the different mode types propagate with different wavelengths, consequently interacting with different parts of the VDF \cite{Verscharen_2019_LRSP}. 
Therefore, the difference in the treatment of $\alpha$ particles has rippling consequences to our understanding of the solar wind dynamics. 
This is the case primarily for the fast solar wind that naturally has higher relative $\alpha$-particle density \cite{Marsch_1982,Kasper_2006_JGRA}. 

%%%%%%%%%%%%%%%%%%%%%%%%
%%% ACKNOWLEDGEMENTS %%%
%%%%%%%%%%%%%%%%%%%%%%%%

\begin{acknowledgments}
Solar Orbiter is a space mission of international collaboration between ESA and NASA, operated by ESA. 
Solar Orbiter SWA data are derived from scientific sensors which have been designed and created, and are operated under funding provided in numerous contracts from the UK Space Agency (UKSA), the UK Science and Technology Facilities Council (STFC), the Agenzia Spaziale Italiana (ASI), the Centre National d’\'Etudes Spatiales (CNES, France), the Centre National de la Recherche Scientifique (CNRS, France), the Czech contribution to the ESA PRODEX programme and NASA. 
The Italian contribution to Solar Orbiter SWA at INAF/IAPS is currently funded under ASI grant 2018-30-HH.1-2022. 
Special thanks are extended to IRAP/CNRS for PAS operations and data calibration, and to the MAG team for providing Solar Orbiter magnetic field data.
M. M. Martinovi\'c and K. G. Klein were financially supported by NASA grants: 80NSSC22K1011, 80NSSC19K1390, 80NSSC23K0693, 80NSSC19K0829, and 80NSSC24K0724.
An allocation of computer time from the UA Research Computing High Performance Computing at the University of Arizona is gratefully acknowledged. D.~Verscharen is supported by STFC Consolidated Grant~ST/W001004/1.
- This research was supported by the International Space Science Institute (ISSI) in Bern, through ISSI International Team project \#563 (Ion Kinetic Instabilities in the Solar Wind in Light of Parker Solar Probe and Solar Orbiter Observations) led by L. Ofman and L.~Jian.
\end{acknowledgments}

%%%%%%%%%%%%%%%%%%%%
%%% BIBLIOGRAPHY %%%
%%%%%%%%%%%%%%%%%%%%


\end{document}